\documentclass[]{spie}  

\usepackage{amsmath,amsfonts,amssymb}
\usepackage{graphicx}
\usepackage[colorlinks=true, allcolors=blue]{hyperref}
\newcommand{\fref}{Fig.~\ref}
\newcommand{\sref}{Sec.~\ref}

\title{Vortex coronagraph: revisiting the phase retrieval properties via Zernike analysis}

\author[a]{G. Orban de Xivry}
\author[a]{O. Absil}

\affil[a]{Space sciences, Technologies, and Astrophysics Research (STAR) Institute, Université de Liège, allée du Six Août 19c, 4000 Liège, Belgium}

\authorinfo{Further author information: (Send correspondence to Gilles Orban de Xivry)\\G. Orban de Xivry: E-mail: gorban@uliege.be.}

\begin{document}
\maketitle

\begin{abstract}
High contrast imaging (HCI) is fundamentally limited by wavefront aberrations, and the ability to perform wavefront sensing from focal plane images is key to reach the full potential of ground and space-based instruments. Vortex focal plane mask coupled with downstream pupil (Lyot) stop stands as one of the best small-angle coronagraphs, but is also sensitive to low-order aberrations. Here, we revisit the behavior of the vortex phase mask, from entrance pupil down to the final detector plane, with Zernike polynomials as input phase aberrations. In particular we develop a second-order expansion that allows us to analyze the phase retrieval properties in a more intuitive and accurate way than previously proposed.
With this formalism, we show how the azimuthal vortex modulation modifies the phase retrieval properties compared to normal imaging. In particular, our results suggest that images obtained with a scalar vortex coronagraph can be used for unambiguous focal-plane wavefront sensing in any practical situation. We compare our results with numerical simulations and discuss practical implementation in coronagraphic instruments.

\end{abstract}


\section{INTRODUCTION}
\label{sec:intro}  

The direct detection of exoplanets relies on coronagraphs to reach the highest contrast. All of them are however limited by the imperfections of the entrance wavefront.
Among the various coronagraph designs, the vortex coronagraph stands out as one of the best small-angle coronagraphs, making it also particularly sensitive to low-order aberrations.
Therefore, enabling  wavefront sensing directly at the focal plane is key to unlock the full potential of future ground and space-based instruments.

\begin{figure}
    \centering
    \includegraphics[width=0.8\linewidth]{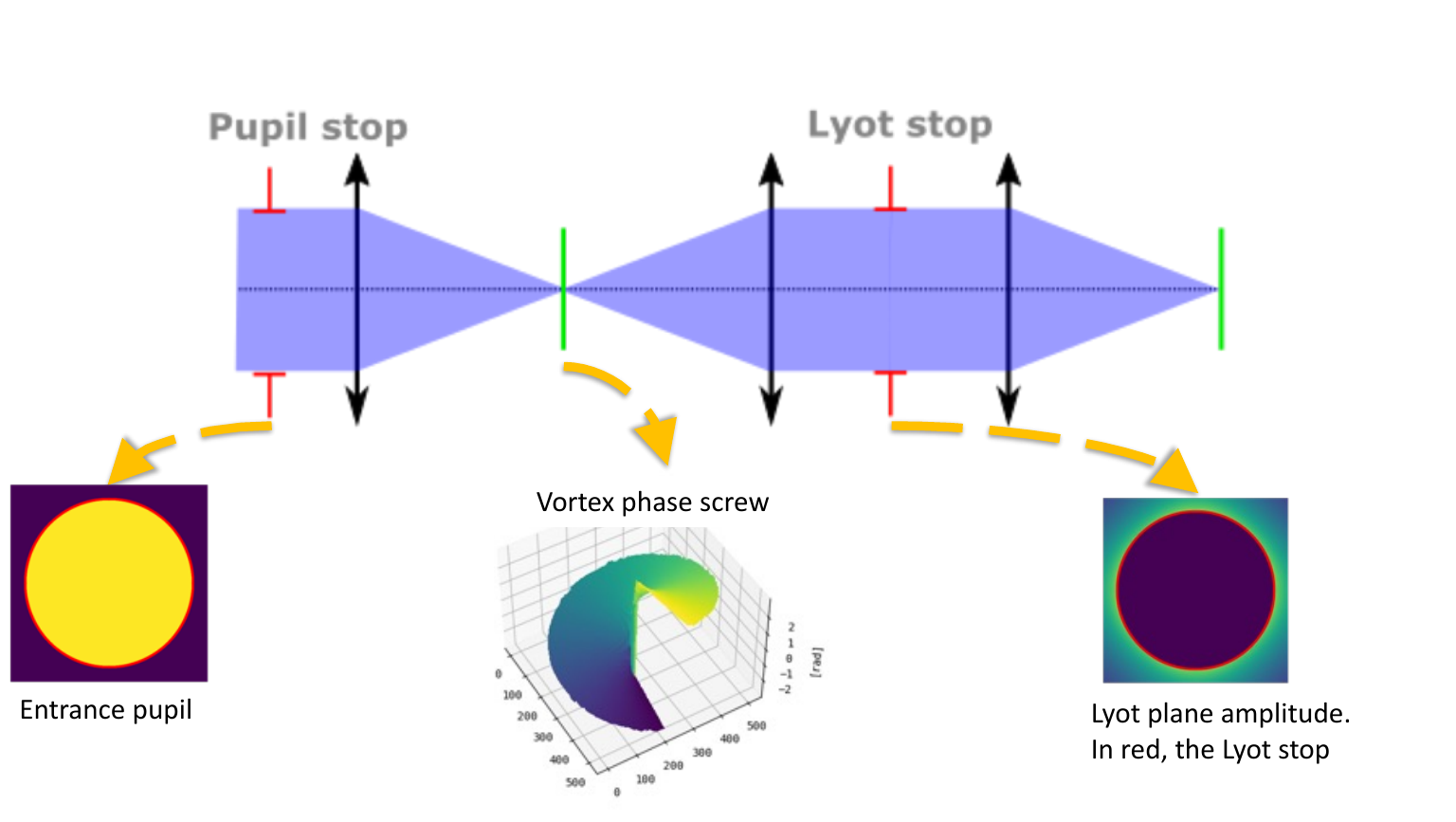}
    \caption{Schematic design of a vortex coronagraph.  From left to right: Entrance circular pupil (red),   intermediate focal plane with the vortex mask (green),  downstream Lyot pupil plane and stop (red), final image plane (green).}
    \label{fig:sketch}
\end{figure}

A vortex coronagraph is composed of a vortex focal plane mask coupled with a downstream pupil (Lyot stop). A schematic layout, \fref{fig:sketch}, illustrates the text book effect of the vortex coronagraph: to move the light of an on-axis source outside the downstream geometric pupil. The diffracted light is then blocked by the Lyot stop, providing theoretically a perfect starlight cancellation for a circular entrance aperture.
The vortex phase mask itself introduces a phase screw $e^{i l_p \phi}$ with  $\phi$ the azimuthal coordinate, and $l_p$ the topological charge of the vortex. There are two flavors of vortex coronagraphs: vectorial (VVC) and scalar (SVC). A vector vortex is based on a half-wave plate with spatially varying fast axis, which results in conjugated phase ramps applied to the two circular polarization states. This is described by the conjugated sign of the topological charge, $\pm | l_p|$.
On the other hand, a scalar vortex  is based on longitudinal phase delay and is described by a single sign of the topological charge. Alternatively, a scalar vortex can be obtained by isolating one circular polarization of a vector vortex.

In order to perform focal-plane wavefront sensing, classical imaging systems typically require a known even aberration diversity to resolve the ambiguity between focal plane images and phase aberrations in the pupil plane.
Here, we study the effect of the vortex azimuthal modulation on the ambiguity between focal plane images and phase aberrations. Specifically, we develop a second-order analytical approximation to describe the electrical field down to the focal plane. This formalism allows us to analyze the phase retrieval properties in an accurate way, and to compare it to classical imaging  and numerical simulations.
In \sref{sec:formalism}, we summarize our analytical approach and provide a few illustrations. In \sref{sec:retrieval}, we use our formalism to analyze the phase retrieval properties of VVC and SVC, first by evaluating the ambiguity on individual Zernike modes and then for the combination of two modes. In \sref{sec:conclusion}, we conclude our contribution.

\section{SECOND-ORDER ZERNIKE EXPANSION FOR THE VORTEX CORONAGRAPH}\label{sec:formalism}

The theoretical expansion we propose is based on three steps. First, we perform a second order expansion of the entrance pupil wavefront in order to express the electrical field as a complex linear combination of Zernike coefficients. Second, we propagate the field from the entrance pupil plane to the downstream Lyot plane, assuming a (perfect) vortex phase mask in the intermediate focal plane. Third, we propagate the field down to the final image plane, where the detector is placed.
The entire formalism is implemented in Symbolic Python (Sympy).

\begin{description}
    \item[Step 1: Pupil plane second order expansion.]
We perform a second order approximation of the entrance pupil wavefront
\begin{align}\label{eq:2nd_pupil}
\begin{split}
E_{\textrm{pup}} &= \exp(i \Phi) \\
                 & \approx 1 + i \Phi - \dfrac{1}{2} \Phi^2\\
                 & \approx 1 + i \sum_{j=2} a_j Z_j - \dfrac{1}{2} \sum_{k=1} b_k Z_k
\end{split}
\end{align}

where the wavefront  is expressed as a combination of Zernike polynomials. The square wavefront $\Phi^2$ is also expressed as a (different) combination of Zernike’s obtained by expressing the products of two Zernike polynomials as a sum of Zernike’s. We use here the result of Marthar\cite{Mathar2009}.

\item[Step 2: Propagation to the Lyot plane.]
After the pupil expansion, we perform the propagation from the pupil to the Lyot plane, assuming a circular aperture and a perfect vortex phase mask of even topological charge $l_p$ in the intermediate focal plane. The linearized expression becomes

\begin{equation}
    E_{\rm Lyot} =  i \sum_{j=2} a_j \zeta_j  - \dfrac{1}{2} \sum_{k=2} b_k \zeta_k
\end{equation}
with $\zeta_j  = \mathcal{F}^{-1}\left[\mathcal{F}\left[ Z_j \right] e^{j l_p \phi} \right]$
denoting the field distribution in the Lyot plane when the input field is defined by the Zernike polynomial $Z_j$. To convert Zernike polynomial $Z_j$ to $\zeta_j$, we use the analytical formalism developed by Huby et al.\cite{Huby+2015}.
Note that we have assumed  a Lyot stop is used, which blocks entirely the diffracted light resulting from the perfect plane component (piston mode). Hence there is no first term, $\zeta_1$, in the equation.

\item[Step 3: Propagation to the image plane.]
The electrical field in the image plane is obtained by taking the analytical Fourier transform of each Zernike polynomial, see for example Niu and Tian\cite{Niu+22}. The final intensity is obtained by taking the square modulus of this linear combination and by evaluating it on a specific grid $(r, \phi)$.

\end{description}

\subsection{Analytical example: astigmatism}
To illustrate the analytical formalism, its differences with the first-order approximation, and the quickly increasing complexity, we calculate here the analytical result for astigmatism alone with an amplitude $a_5$.
In the following, we assume a topological charge $l_p=2$.

The linear approximation provides the following results
\begin{equation}
E_{\rm pup}\approx 1 + i a_5 Z_5
\longrightarrow
E_{\rm lyot} = - a_5 \frac{\sqrt{2} Z_4}{2}
\longrightarrow
E_{\rm img}=a_5 \frac{\sqrt{6} J_{3}\left(2 \pi k\right)}{4 \pi k}
\end{equation}

The resulting electrical field in the image plane is a pure radial function composed of a single Bessel function. Consequently, this implies that both positive and negative signs of the astigmatism would produce two identical intensity images.\\

The second order approximation provides the following results

\begin{align}
E_{\textrm{pup}} &=
(1 - \frac{a_5^2}{2}) + i a_5 Z_5  + a_5^2 \left(-  0.25 \sqrt{3} Z_4  -  0.05 \sqrt{5} Z_{11} +  0.15 \sqrt{10} Z_{14} \right)\\
E_{\textrm{lyot}} &=  - a_5 \frac{\sqrt{2} Z_4}{2} \nonumber \\
  & \mathrel{\phantom{=}}+ a_5^2 \left\{
     -  0.05 \sqrt{5} \left(\frac{\sqrt{2} Z_{12}}{2} + \frac{\sqrt{2} i Z_{13}}{2}\right)
     - 0.25 \sqrt{3} \left(\frac{\sqrt{2} i Z_5}{2} + \frac{\sqrt{2} Z_6}{2}\right) \right.\\
   &\mathrel{\phantom{= }} \left.
     + 0.15 \sqrt{10} \left(0.5 Z_{12} - \frac{i Z_{13}}{2}\right)
     \right\} \nonumber \\
E_{\textrm{img}} &= a_5 \frac{\sqrt{6} J_{3}\left(2 \pi k\right)}{4 \pi k} \nonumber \\
   & \mathrel{\phantom{=}} + a_5^2 \left\{- \frac{0.375 i \sin{\left(2 \phi \right)} J_{3}\left(2 \pi k\right)}{\pi k} + \frac{0.5 i \sin{\left(2 \phi \right)} J_{5}\left(2 \pi k\right)}{\pi k} \right. \\
    & \mathrel{\phantom{= }} \left.
    + \frac{0.375 \cos{\left(2 \phi \right)} J_{3}\left(2 \pi k\right)}{\pi k}
    + \frac{0.25 \cos{\left(2 \phi \right)} J_{5}\left(2 \pi k\right)}{\pi k} \right\} \nonumber
\end{align}

The resulting field in the image plane is no longer purely radial, and  the images differ depending on whether the sign of astigmatism is positive or negative.

\subsection{Illustration: comparing first order and second-order approximation}

We illustrate below the better representativeness of the second order approximation.
The difference between first and second order was already observed for tip-tilt by Huby et al\cite{Huby+2015}.
Here we generalize the second order formalism to all Zernike’s and investigate the phase retrieval properties.

\begin{figure}[ht]
    \includegraphics[width=0.33\linewidth]{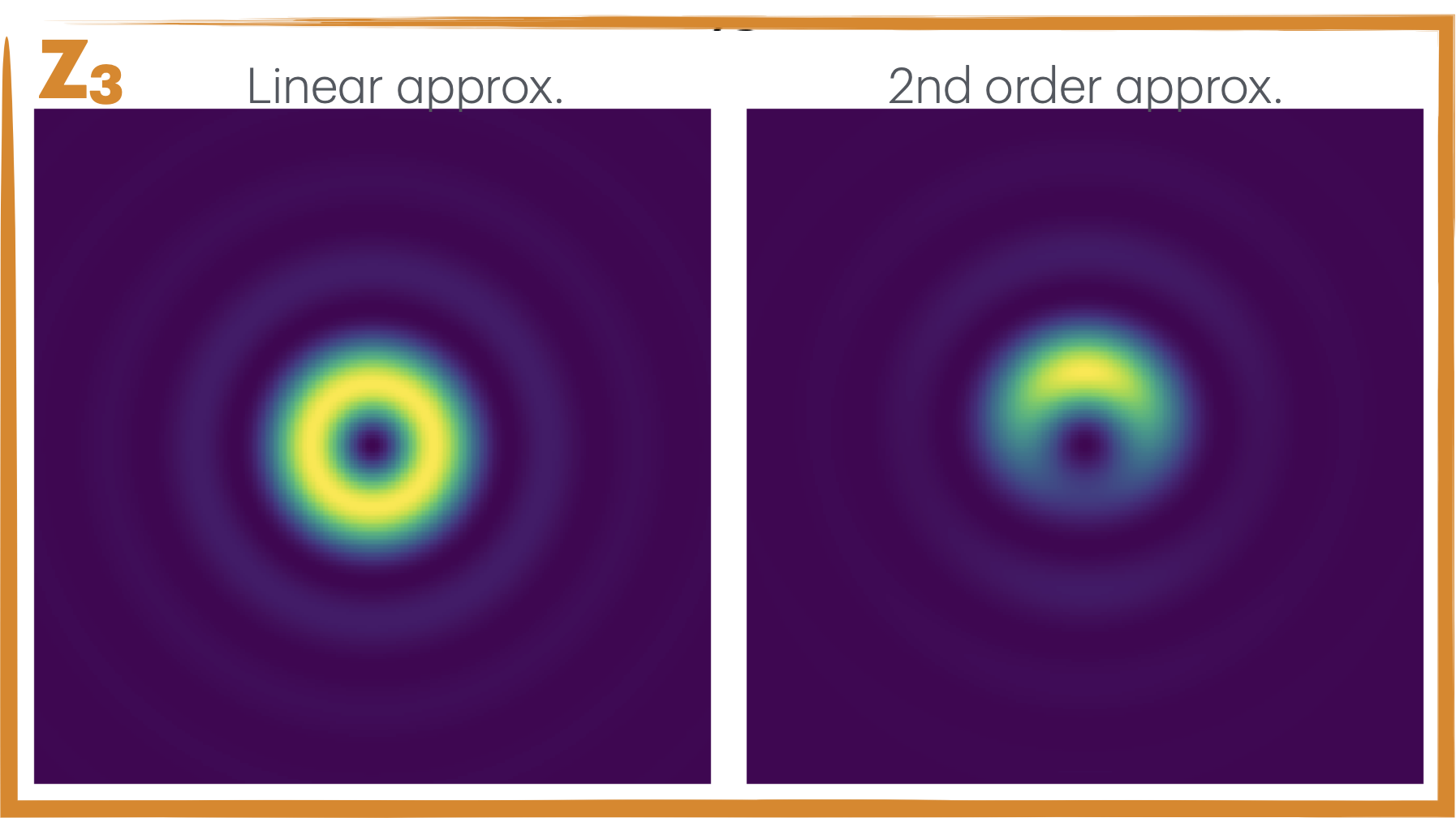}
    \includegraphics[width=0.33\linewidth]{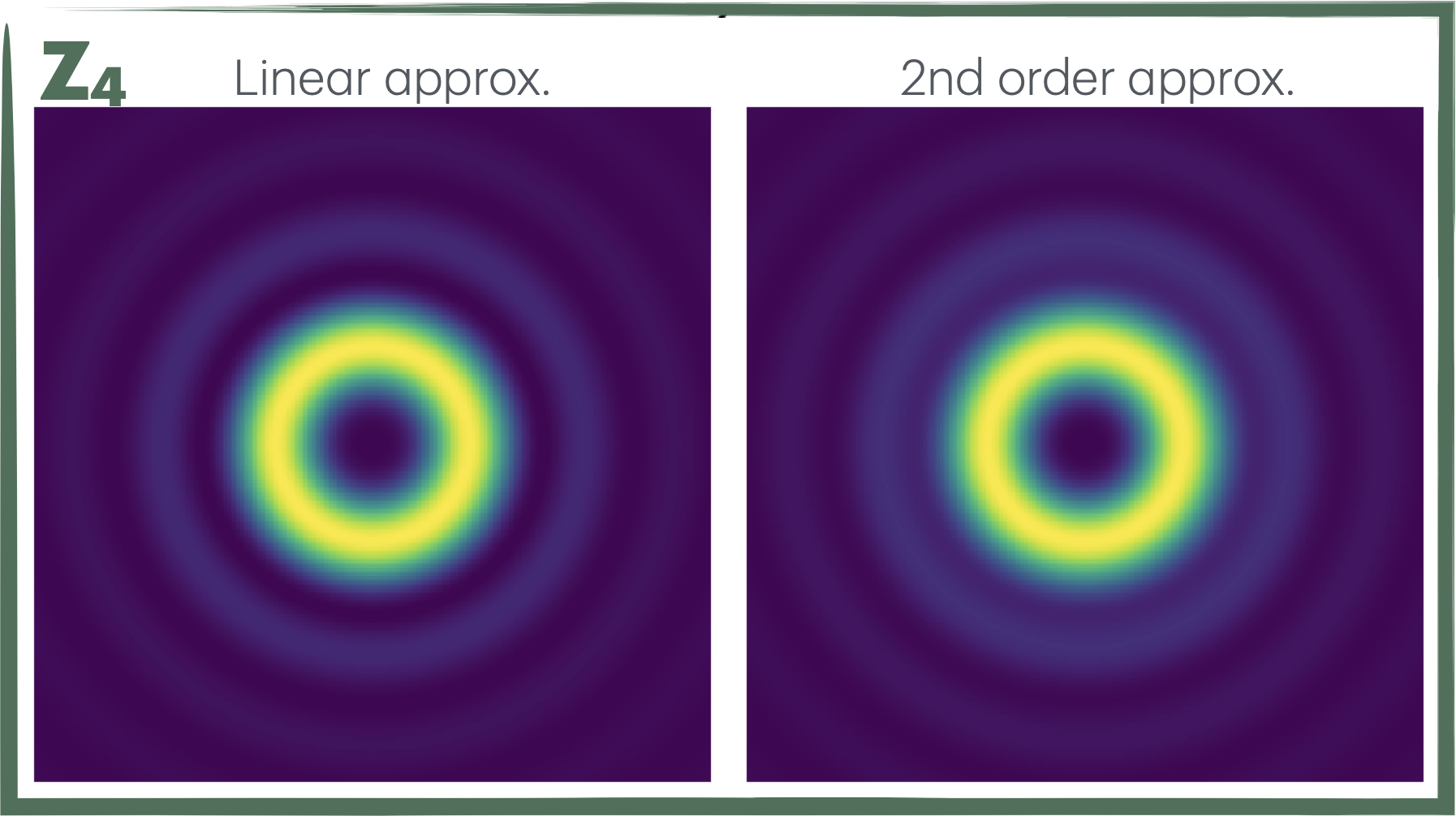}
    \includegraphics[width=0.33\linewidth]{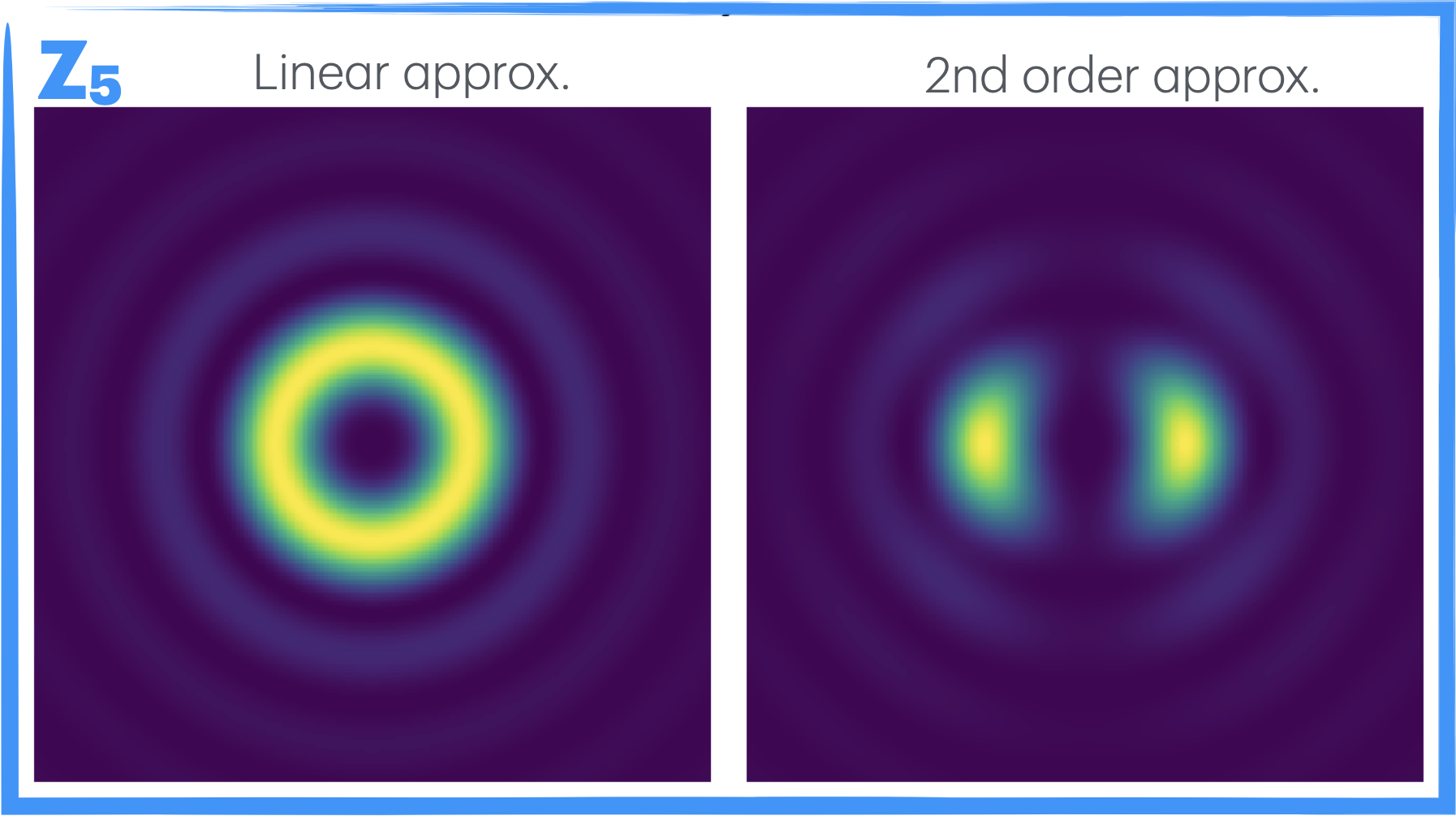}
    \caption{Comparison between first and ssecond order approximation for tilt, focus, astigmatism ($Z_3$, $Z_4$, $Z_5$). In each case we use of scalar vortex of topological charge $l_p=2$, and an aberration level of 1~rad rms.}
    \label{fig:enter-label}
\end{figure}

The second order approximation reveals the asymmetry of the tilt ($Z_3$) and the astigmatism ($Z_5$). An opposite sign for the astigmatism leads to a rotation of 90 degrees of the pattern.
Focus remains a purely radial function, which can also be verified at the third order.

\subsection{Comparison with numerical approaches}

It is interesting to compare our second order approach with numerical simulations to establish the validity range of our approximation, and understand any numerical artifacts.

A numerical representation of a pure vortex screw on a finite grid is challenging, especially close to the singularity (the center) where the phase rapidly changes over a limited number of pixels. At least two approaches are typically considered, individually or in combination, to obtain an accurate representation of the phase ramp: oversampling the central region, and numerically enforcing the perfect cancellation in the downstream pupil.

The first method involves computing the (central) phase ramp on a larger grid and then rebinning the complex array down to the normal sampling of the simulation. The method produces good results but not a perfect representation of the vortex phase ramp.
The second method allows to free the numerical simulation from any aliasing and Fourier transform artifacts. Specifically, it ensures that the perfect vortex with an unaberrated field is perfectly simulated according to theory, i.e., the intensity inside the downstream (Lyot) pupil plane is zero for a circular pupil. To enforce this, the perfect field plus perfect coronagraph are propagated to the Lyot stop, then the intensity is zeroed-out inside the pupil and back-propagated to the focal plane. The resulting electrical field can then be used for future propagations.
In the following, we use a custom numerical simulator, which only implements the second method (back-propagation), but not the oversampling.

\begin{figure}[t]
    \includegraphics[width=1.\linewidth]{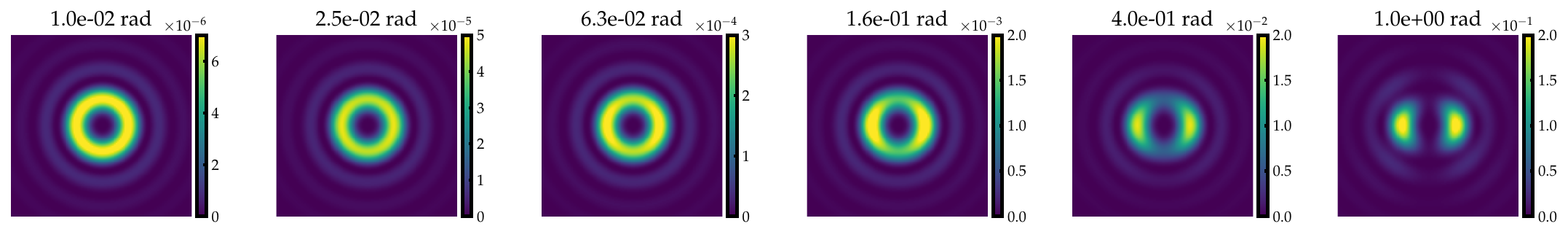}
    \includegraphics[width=\linewidth]{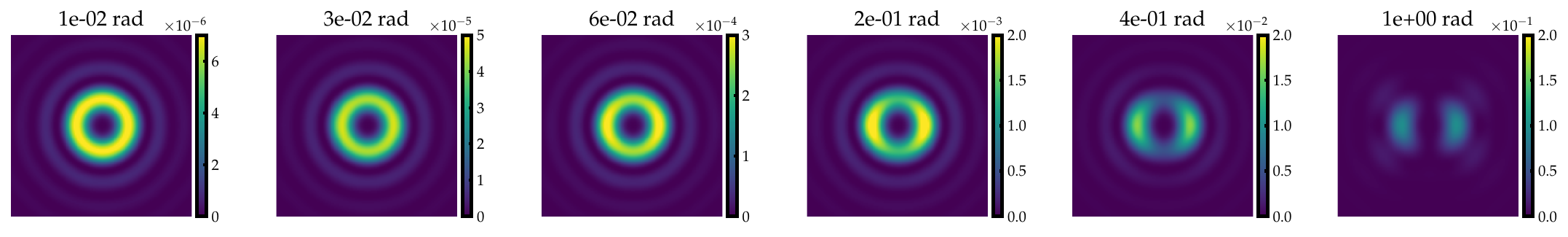}
    \caption{Vortex coronagraphic images comparing analytical (top) and numerical simulations for different level of astigmatism between 1e-2 and 1 rad rms.}
    \label{fig:illustration}
\end{figure}

We perform a first visual comparison  by injecting astigmatism with increasing amplitudes up to 1 rad rms, see \fref{fig:illustration}. The images are normalized by their off-axis PSF, which allows a direct comparison between the two approaches and informs us on the raw contrast. The morphology and the flux of the PSFs  match closely but begin to diverge at around 0.4 rad, reaching a factor of more than 2 in intensity at around 1 rad. This indicates that the second-order analytical approach becomes invalid.

\begin{figure}[t]
    \centering
    \includegraphics[width=0.6\linewidth]{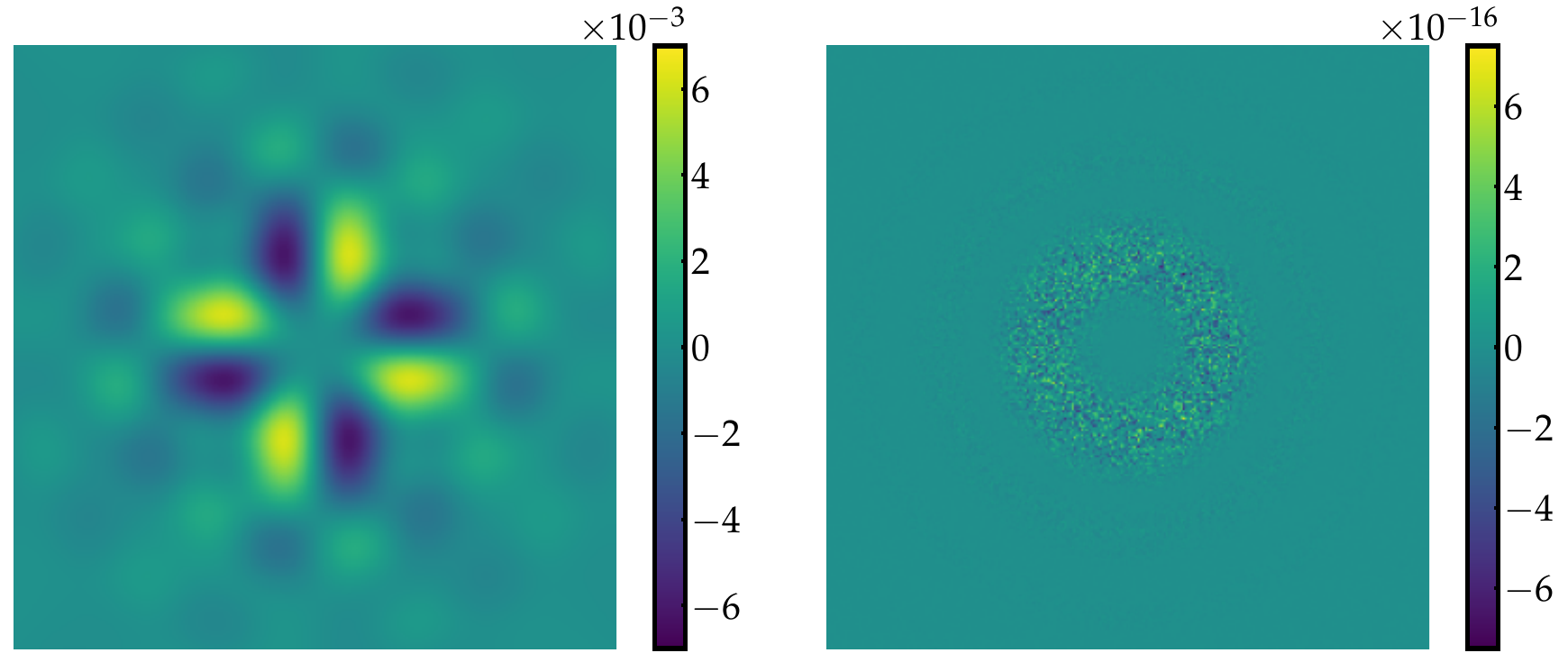}
    \caption{Difference between images with opposite sign of focus. (Left) based on numerical simulations of a charge-2 SVC. (Right) based on our second order analytical approach.}
    \label{fig:focus}
\end{figure}

We now turn our attention to a pure radial mode: focus.
We numerically simulate the propagation of a small defocus through an SVC of charge 2 and compare the images obtained for opposite signs of the mode. We perform the same using the analytical approach.
The differential images, normalized by the peak intensity of one propagation, are illustrated in \fref{fig:focus}, with a 4$\lambda/D$ radial field-of-view and 32 pixels per $\lambda/D$.
While the difference for the analytical approach leads to random numerical noise only, the Fourier-based simulations lead to a distinctive pattern which would imply that there is no sign ambiguity on this mode. This is a numerical artifact due to the finite sampling  of the vortex phase ramp, and the entrance and Lyot pupils.
Unsurprisingly, we expect that the sampling precision, especially in the central region of the vortex screw, will always limit the fidelity of vortex coronagraphic simulations to some extent and require dedicated attention.

\section{SIGN AMBIGUITIES WITH VORTEX CORONAGRAPHS}\label{sec:retrieval}
\subsection{Individual Zernike modes}
For circularly symmetric pupils and classical imaging, there is no unique relationship between focal plane images
and phase aberrations in the pupil plane. For point-like sources, the non-uniqueness consists in a (single) sign
ambiguity on even modes.

Based on this consideration, to evaluate the ambiguity between two images (obtained with classical or coronagraphic imaging) with phase aberrations $\phi_1$ and $\phi_2$, we use a simple L2 sensitivity metric based on the final images. A null value means the images are identical. For non-zero values, we can expect that the higher the better to perform phase retrieval.
We compute the sensitivity for the first 13 Zernike modes starting from tilt ($Z_3$), using an amplitude of 0.1~rad. We perform the computation both for a scalar and a vector vortex
of topological charge 2, as well as for non-coronagraphic imaging as a reference.
Similarly to normal imaging,
we can see that the vector vortex is insensitive to the sign of even modes. The scalar vortex displays a different
behavior, where only the pure radial modes are ambiguous (here focus $Z_4$ and spherical aberration $Z_{11}$), and
the other even modes are unambiguous (here, oblique astigmatism $Z_5$ and vertical astigmatism $Z_6$, but also $Z_{12}$, $Z_{13}$, $Z_{14}$, and $Z_{15}$).
For non-even modes, SVC and VVC display the same sensitivity with some slight discrepancies from classical
imaging.

\begin{figure}[p]
    \centering
    \includegraphics[width=\linewidth]{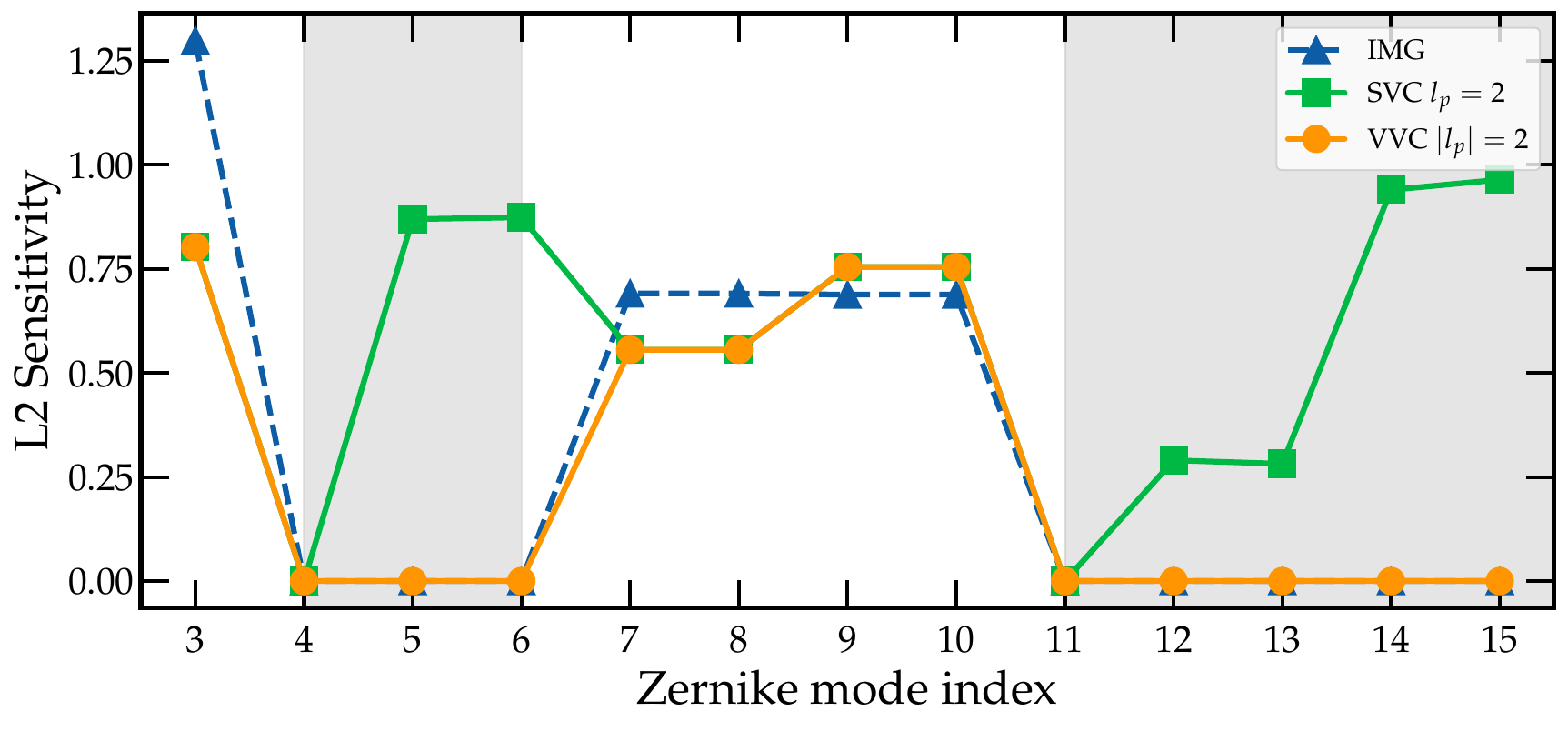}
    \caption{Sensitivity to individual modes, from tilt ($Z_3$) to $Z_{15}$. The grey shaded area denotes even Zernike modes. $Z_4$ (focus) and $Z_{11}$ (spherical) are pure radial modes.}
    \label{fig:enter-label}
\end{figure}

This phase retrieval property ensues from the azimuthal modulation
as opposed for example to the classical focus phase diversity, which introduces a radial modulation. It follows
that pure radial modes (e.g., focus and spherical) are not modulated by the phase screw in an informative way and the sign ambiguity cannot be lifted.

\subsection{Combination of Zernike modes}
\begin{figure}[p]
    \centering
    \includegraphics[width=0.7\linewidth]{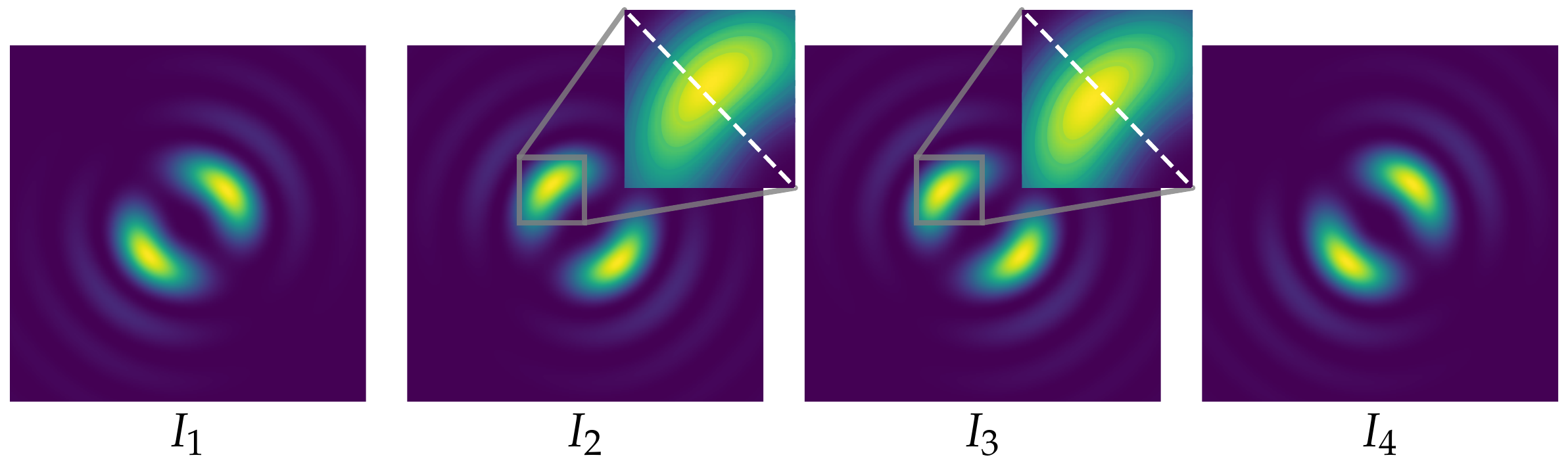}
    \caption{(Left) Four images obtained with a combination of $Z_4$ and $Z_5$ with equal rms of $0.5/\sqrt{2}$. The zoom insets emphasize the subtle difference. }
    \label{fig:focus+asti}
\end{figure}

\begin{figure}[p]
    \centering
    \includegraphics[width=0.5\linewidth]{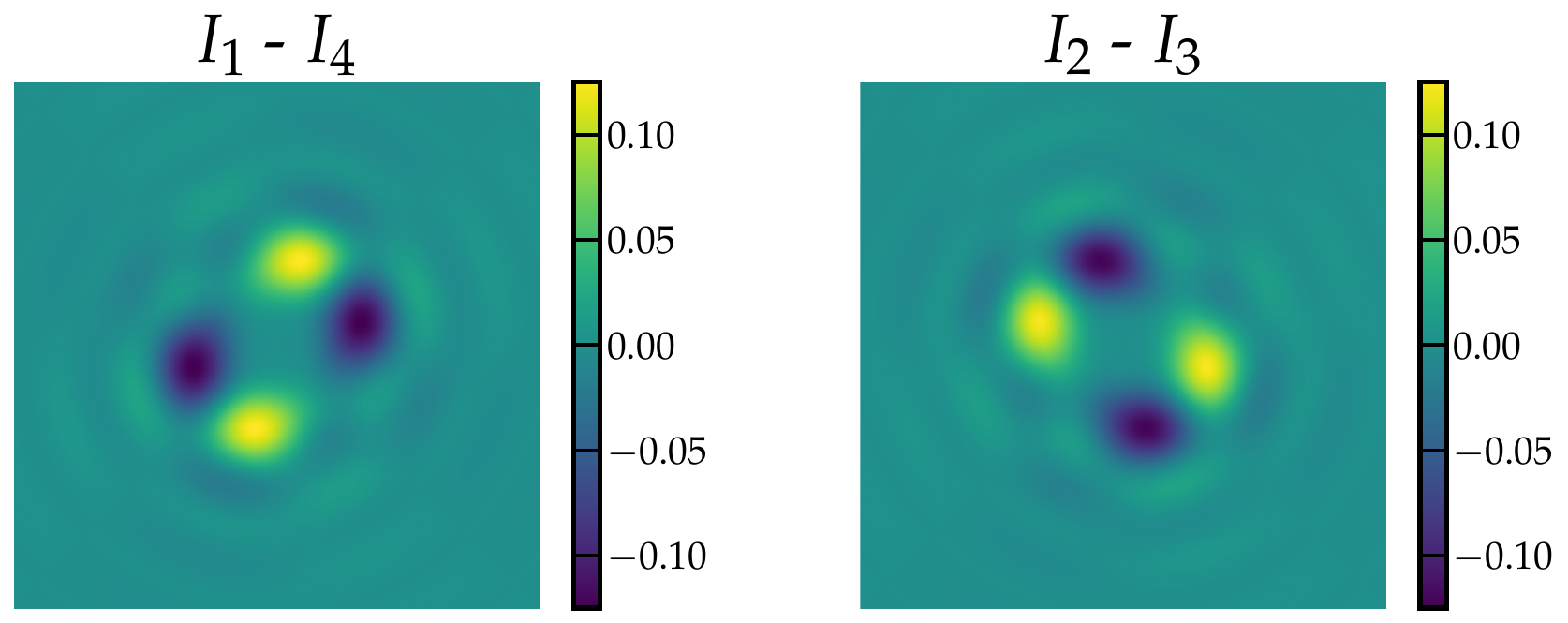}
    \caption{Subtraction of the two close pairs, $(I_1-I_4)$ and $(I_2 –I_3)$, illustrated in \fref{fig:focus+asti}.}
    \label{fig:diff}
\end{figure}

With the azimuthal modulation introduced by the vortex, the ambiguity on even modes breaks down and we can anticipate that with a combination of modes the ambiguity on pure radial modes may be lifted.
We consider a combination of focus and astigmatism with equal amplitudes ($0.5/\sqrt{2}$ rad rms each). By considering all sign combinations of those two modes,  we obtain 4 different images, see \fref{fig:focus+asti}.
Two pairs are closely similar but not identical, as  observed in the zoom insets and  illustrated by the image difference \fref{fig:diff}.
From these illustrations, we can consider the ambiguity lifted, albeit with presumably lower sensitivity.
Arguably, this conclusion can be extended to any combination of modes, suggesting that sufficient information exists to lift sign ambiguities in the focal plane image of a scalar
vortex coronagraph.

\section{CONCLUSION}\label{sec:conclusion}

In conclusion, our results suggest that phase aberrations can be  retrieved in any practical scenario using scalar vortex coronagraph (SVC) images.
This finding corroborates the work of Riaud et al.\cite{Riaud+2012, Riaud+2012b}, and is further supported by Quesnel et al.\cite{Quesnel+2022} where
deep learning techniques were specifically employed to successfully retrieve phase aberrations based on vortex coronagraphic images.

We also note that numerical simulations can  be misleading by generating non random numerical artifacts: they may, for example, artificially resolve ambiguities in pure radial modes, which is theoretically not expected. In this respect, our analytical approach may provide a more reliable ground-truth in the low aberration regime.

Real-world systems introduce additional complexities, including more intricate pupil designs, the use of downsized Lyot stops, and the inherent limitations of the vortex coronagraph, such as the leakage term. These factors can potentially constrain the accuracy of phase retrieval, and should be considered when applying phase retrieval techniques in high contrast imaging systems with vortex coronagraphs.

\bibliography{report} 
\bibliographystyle{spiebib} 
\end{document}